# Giant exchange bias in a single-phase magnet with two magnetic sublattices

Y. Sun,[1,a)] J.-Z. Cong,[1] Y.-S. Chai,[1] L.-Q. Yan,[1] Y.-L. Zhao,[1] S.-G. Wang,[1] W. Ning,[2] and Y.-H Zhang[2]

[1]Beijing National Laboratory for Condensed Matter Physics, Institute of Physics, Chinese Academy of Sciences, Beijing 100190, China

[2]High Magnetic Field Laboratory, Chinese Academy of Sciences, Hefei, Anhui 230031, China

## Abstract

Exchange bias phenomenon is generally ascribed to the exchange coupling at the interfaces between ferromagnetic and antiferromagnetic layers. Here we propose a bulk form of exchange bias in a single-phase magnet where the coupling between two magnetic sublattices induces a significant shift of the coercive field after a field cooling. Our experiments in a complicated magnet $YbFe_2O_4$ demonstrate a giant exchange bias at low temperature when the coupling between the $Yb^{3+}$ and $Fe^{2+}/Fe^{3+}$ sublattices take places. The cooling magnetic field dependence and the training effect of exchange bias are consistent with our model. In strong contrast to conventional interfacial exchange bias, this bulk form of exchange bias can be huge, reaching the order of a few Tesla.

[a)]Author to whom correspondence should be addressed. Electronic mail: youngsun@iphy.ac.cn

Exchange bias (EB) was initially observed in Co particles coated with a layer of CoO by Meiklejohn and Bean.[1] It refers to the shift of magnetic hysteresis loop along the field axis after the sample is cooled in a magnetic field. Since then, EB has been found in a large number of heterostructures such as core-shell nanoparticles, magnetic multilayers and thin films.[2-5] Compared with heterogeneous structures, EB is rare in structurally single-phased bulk materials. Recently, there were a few reports of EB in bulk compounds with intrinsic phase separation[6-8] or spin glass phase.[9-11] In general, these EB phenomena are ascribed to the exchange coupling at different magnetic interfaces that causes a pinning of magnetization. In this Letter, we propose a bulk form of EB where the pinning phenomenon is caused by the global interaction between two magnetic sublattices instead of the interfacial exchange coupling.

As illustrated in Fig. 1(a), we consider a single-phase magnet that consists of two magnetic sublattices − one with antiferromagnetic (AFM) structure and another with ferromagnetic (FM) structure. The spin configuration is therefore similar to that of an artificial FM/AFM superlattice. In the simplest case, we assume that the easy axis of the AFM and FM sublattices are in the same direction along the c axis. The energy density of the system can be expressed as

$$E/V = -\vec{H}\cdot\vec{M} + K_{FM}\sin^2(\alpha) + K_{AFM}\sin^2(\beta) - J_{INT}\cos(\theta), \qquad (1)$$

where $\vec{H}$ is the applied magnetic field, $\vec{M}$ is the magnetization of the system, $K_{FM}$ and $K_{AFM}$ are the magnetic anisotropy constants of the FM and AFM sublattices, and $J_{INT}$ is the exchange coupling constant between the two sublattices. The angles $\alpha$ and $\beta$ are the angles between the spins and easy axis in the FM and AFM subatttices, respectively. $\theta$ is the angle between the FM and AFM spins. After a magnetic-field cooling (FC) process from a high temperature ($T > T_N$, $T_C$) to a low temperature ($T < T_N$, $T_C$), the system is driven into a uniform spin configuration like Fig.

1(a). If there were a strong coupling between two sublattices, $J_{INT} > K_{AFM} > K_{FM}$, the AFM and FM spins would rotate together and no EB appears. However, in the case of a large magnetic anisotropy of the AFM sublattice and a moderate inter-sublattice coupling, $K_{AFM} > J_{INT}$ and $K_{AFM} > K_{FM}$, the FM spins rotates with external magnetic fields whereas the AFM spins remain the original configuration due to the large $K_{AFM}$. Similar to that occurred at a FM/AFM interface, the spin rotation of the FM sublattice feels the pinning force from the AFM sublattice due to $J_{INT}$. As a result, a shift of the magnetic hysteresis loop along the field axis could be expected.

We have testified this model in a complicated magnet $YbFe_2O_4$ in which the pinning of magnetization is due to the interplay between $Yb^{3+}$ ($4f^{13}$; $4.5\mu_B$) and $Fe^{2+}/Fe^{3+}$ magnetic sublattices. A giant EB effect is indeed observed at low temperature as expected. To elaborate our model more convincingly, $LuFe_2O_4$ is chosen as a compare. Since $Lu^{3+}$ ($4f^{14}$, $0\mu_B$) is a non-magnetic ion, the absence of interplay between magnetic sublattices should not give rise to the proposed EB, which is consistent with our experimental results.

$YbFe_2O_4$ and $LuFe_2O_4$ belong to the family of $RFe_2O_4$ (R=Y, Ho, Er, Tm, Yb and Lu). All of them have a hexagonal layered structure. As shown in Fig. 1(b), the Fe-O triangular lattices bilayer and the R-O layer alternatively stack along c axis. In the Fe-O bilayer structure, there are equal numbers of $Fe^{2+}$ and $Fe^{3+}$ ions. The magnetic properties of $LuFe_2O_4$ have been intensively studied in the past decade.[12-15] Below ~ 240 K, the $Fe^{2+}/Fe^{3+}$ moments form a ferrimagnetic (FI) ordering with the easy axis along c axis.[15] Because of the layered structure and a large unquenched orbital magnetic moment, $LuFe_2O_4$ exhibits a strong magnetic anisotropy with a huge coercivity.[16,17] While they have the same crystal structure, the large magnetic moment (4.5 $\mu_B$) of $Yb^{3+}$ ion makes the magnetic behaviors of $YbFe_2O_4$ more fascinating than $LuFe_2O_4$. In this work, we demonstrate that the interplay between the $Yb^{3+}$ and $Fe^{2+}/Fe^{3+}$ sublattices can induce a giant EB effect.

$YbFe_2O_4$ and $LuFe_2O_4$ single crystals were prepared by the method of optical floating-zone melting in a flowing argon atmosphere. X-ray diffraction (XRD) and back-reflection Laue XRD experiments were taken to check the crystallization and determine the crystallographic direction. Powder XRD measurements at room temperature and Rietveld analysis indicate that the samples are single phase and have a structure consistent with literature. The temperature dependence of magnetization was measured with a Quantum Design superconducting interference device (SQUID) magnetometer, and the *M-H* hysteresis loops up to 13 T were measured with a Quantum Design physical property measurement system (PPMS). All the magnetization measurements were performed along the crystallographic *c* axis which is the magnetic easy axis.

Figure 2 presents the temperature dependence of magnetization for $YbFe_2O_4$ in a low magnetic field (1 kOe) with the FC mode. There are apparently two magnetic transitions. The ferrimagnetic ordering of $Fe^{2+}/Fe^{3+}$ magnetic moments is founded at $T_1 \sim 245$ K, similar to that in $LuFe_2O_4$. There is another magnetic transition around 50 K ($T_2$) due to the ordering of $Yb^{3+}$ magnetic moments. Below 50 K, the FC magnetization drops with decreasing temperature, which indicates that the spontaneous interaction between $Yb^{3+}$ and $Fe^{2+}/Fe^{3+}$ sublattices prefer a canted AFM configuration. What makes the system interesting is that the inter-sublattice coupling is relatively weak and can be manipulated by a FC process under moderate magnetic fields. As shown in the inset of Fig. 2, after cooling the sample in 5 or 13 T from 300 K, the FC magnetization goes up instead of dropping down below $\sim 50$ K ($T_2$), indicating a parallel alignment of two magnetic sublattices. Thus, the relative orientation between $Yb^{3+}$ and $Fe^{2+}/Fe^{3+}$ magnetic sublattices can be effectively manipulated by the FC process.

Figure 3 shows the *M-H* hysteresis loops at 5 K for $YbFe_2O_4$ with both the zero field cooling (ZFC) and FC modes. In the FC mode, $YbFe_2O_4$ was cooled from 300 to 5 K with a cooling field $H_{cool}$=13 T. While the ZFC *M-H* loop exhibits nearly symmetric coercivity, the FC *M-H* loop shifted left along *H* axis with $H_1$=−71.2 kOe and $H_2$=39.3 kOe, where $H_1$ and $H_2$ is the left and right coercive field, respectively. The EB field ($H_{EB}$), defined as $H_{EB}$=−$(H_1+H_2)/2$, is 15.8 kOe. Such a value of $H_{EB}$ is much larger than those reported ever before, and is really a giant EB effect. In order to prove that the interplay between two magnetic sublattices is the source of the observed giant EB, we measured the *M-H* loop of $LuFe_2O_4$ as a compare because $Lu^{3+}$ is a nonmagnetic ion. As shown in the inset of Fig. 3, the FC *M-H* hysteresis loop of $LuFe_2O_4$ measured in the same condition is nearly symmetric with $H_1$= −95.0 kOe and $H_2$= 94.8 kOe. This huge coercive field suggests a large magnetic anisotropy constant of the $Fe^{2+}/Fe^{3+}$ sublattice in consistence with previous reports.[16] The absence of EB in $LuFe_2O_4$ confirms that the giant EB is related to the interplay between $Yb^{3+}$ and $Fe^{2+}/Fe^{3+}$ magnetic sublattices.

We then studied how the EB depends on the cooling magnetic field $H_{cool}$ at 5 K. The sample was cooled from 300 to 5 K under different $H_{cool}$=5, 10, 30, 50, and 130 kOe. When the temperature is stabilized at 5 K, the magnetic field was set to 130 kOe and the *M-H* loops were measured between ±130 kOe. As shown in Fig. 4, $H_{EB}$ increases rapidly with increasing $H_{cool}$ and reaches a maximum of 19.6 kOe at $H_{cool}$=10 kOe. With further increasing cooling field, $H_{EB}$ decreases slightly and remains a high level up to 13 T. This feature of cooling field dependence is distinct from that in the phase-separated oxides or spin glassy materials where $H_{EB}$ decays rapidly with high cooling magnetic fields.[5,6,18,19] This distinction suggests that the observed giant EB is not related to phase separation or spin glassy phase.

In order to further confirm the giant EB effect, we then studied the temperature dependence of EB. As plotted in Fig. 5, $H_{EB}$ decreases rapidly with increasing temperature, and becomes negligible above ~ 50 K. At 25 K, the FC hysteresis loop still shows a clear shift (inset of Fig. 5), though the $H_{EB}$ is reduced from ~ 19 kOe at 5 K to ~ 3 kOe at 25 K. The cooling field dependence of $H_{EB}$ at 25 K exhibits a similar trend to that at 5 K, increasing fast in low cooling field and slightly decreasing in high cooling fields. The temperature dependence of $H_{EB}$ proves that the EB effect appears only after the interaction between $Yb^{3+}$ and $Fe^{2+}/Fe^{3+}$ magnetic sublattices develops below $T_2$ (~ 50 K).

The above experiments in $YbFe_2O_4$ support our model of a bulk form of EB induced by the interplay between two magnetic sublattices. Though the spin configuration of $YbFe_2O_4$ is not exactly identical to the ideal case illustrated in Fig. 1(a), it holds two important features required by our model. Firstly, the coupling between two magnetic sublattices ($Yb^{3+}$ and $Fe^{2+}/Fe^{3+}$) is moderate and can be modified by the FC process. Secondly, the FI sublattice ($Fe^{2+}/Fe^{3+}$) has a large magnetic anisotropy constant so that it can act effectively as the AFM component in our model.

Another important property of EB is the so-called training effect,[20,21] *i.e.*, the EB decreases monotonically with the cycling number (*n*) of consecutive hysteresis loops. It has been known that there are two types of training effect,[22] one between the first and second loop and another involving subsequent higher number of loops ($n \geq 2$). In the FM/AFM thin films, the first type of training effect has been proposed to arise from the AFM magnetic symmetry.[23] For the second type of training effect, it is often found experimentally that the relationship between $H_{EB}$ and the cycling number $n$ follows a simple power law

$$H_{EB}(n) - H_{EB}(\infty) = \kappa / \sqrt{n}, \qquad (2)$$

where $\kappa$ is a system-dependent constant and $H_{EB}(\infty)$ is the exchange-bias field in the limit of infinite loops.[24]

Figure 6 presents the training effect in YbFe$_2$O$_4$. After FC from 300 to 5 K with $H_{cool}$=13 T, the $M-H$ loops were measuredly continuously in the range of ±13 T up to 10 cycles. The gradual decrease of $H_{EB}$ with increasing loop number $n$ is clearly seen, evidencing a training effect of EB. As shown in the inset of Fig. 6, except a strong decay between the first and the second loop, $H_{EB}(n)$ follows the power law of Eq. (2) for n ≥ 2, with the parameters $\kappa$=14.5 kOe and $H_{EB}(\infty)$=2.6 kOe. In FM/AFM heterostructures, the training effect in general has its origin from the rearrangement of the AFM domain structure at the FM/AFM interface which takes place during each magnetization reversal of the FM layer.[21,25] In other words, the strength of the training effect depends significantly on the properties of the AFM pinning layer rather than the FM layer. Similarly, the training effect in YbFe$_2$O$_4$ indicates that the spin structure of the system deviates from its equilibrium configuration and relaxes during each magnetization reversal. This relaxation of spin configuration occurs mainly in the FI sublattice (Fe$^{2+}$/Fe$^{3+}$) that acts as the AFM pinning layer. The coupling between two magnetic sublattices brings two sides of one coin. On one side, it causes a unidirectional pinning of magnetization to the FM sublattice that induces the EB effect. On another hand, it triggers the relaxation and reorientation of domains in the AFM or FI sublattice during the magnetization reversal process that gives rise to the training effect.

In summary, we have proposed a bulk form of EB in a single-phase magnet with two interacting magnetic sublattices. Comparative experiments between the isostructural $LuFe_2O_4$ and $YbFe_2O_4$ confirm a giant EB in the order of a few Tesla induced by the interplay between $Yb^{3+}$ and $Fe^{2+}/Fe^{3+}$ magnetic sublattices. This bulk form of EB also exhibits a training effect similar to that in conventional interfacial EB systems. We may expect this type of EB in other complicated magnets with properly interacting magnetic sublattices and strong magnetic anisotropies.

This work was supported by the Natural Science Foundation of China under Grant Nos. 11074293, 51021061, and 11104337 and the National Key Basic Research Program of China under Grant No. 2011CB921801.

References

[1]W. P. Meiklejohn and C. P. Bean, Phys. Rev. **102**, 1413 (1956).

[2]J. Nogués and Ivan K. Schuller, J. Magn. Magn. Mater. **192**, 203 (1999).

[3]M. Kiwi, J. Magn. Magn. Mater. **234**, 584 (2001).

[4]J. Nogués, J. Sort, V. Langlais, V. Skumryev, S. Suriñach, J. S. Muñoz, and M. D. Baró, Phys. Rep. **422**, 65 (2005).

[5]S. Giri, M. Patra, and S. Majumdar, J. Phys.: Condens. Matter **23**, 073201 (2011).

[6]D. Niebieskikwiat and M. B. Salamon, Phys. Rev. B **72**, 174422 (2005).

[7]Y. K. Tang, Y. Sun, and Z. H. Cheng, Phys. Rev. B, **73**, 174419 (2006).

[8]L. Pi, S. Zhang, S. Tan, and Y. Zhang, Appl. Phys. Lett. **88**, 102502 (2006).

[9]J. S. Kouvel, and W. Abdul-Razzaq, J. Magn. Magn. Mater. **53**, 139 (1985).

[10]L. Ma, W. H. Wang, J. B. Lu, J. Q. Li, C. M. Zhen, D. L. Hou, and G. H. Wu, Appl. Phys. Lett. **99**, 182507 (2011).

[11]L.-Q. Yan, F. Wang, Y. Zhao, T. Zou, J. Shen, and Y. Sun, J. Magn. Magn. Mater. **324**, 2579 (2012).

[12]A. Nagano, M. Naka, J. Nasu, and S. Ishihara, Phys. Rev. Lett. **99**, 217202 (2007).

[13]A. D. Christianson, M. D. Lumsden, M. Angst, Z. Yamani, W. Tian, R. Jin, E. A. Payzant, S. E. Nagler, B. C. Sales, and D. Mandrus, Phys. Rev. Lett. **100**, 107601 (2008).

[14]C. H. Li, F. Wang, Y. Liu, X. Q. Zhang, Z. H. Cheng, and Y. Sun, Phys. Rev. B **79**, 172412 (2009).

[15]K. Kuepper, M. Raekers, C. Taubitz, M. Prinz, C. Derks, M. Neumann, A. V. Postnikov, F. M.

F. de Groot, C. Piamonteze, D. Prabhakaran, and S. J. Blundell, Phys. Rev. B **80**, 220409(R) (2009).

[16]W. Wu, V. Kiryukhin, H.-J. Noh, K.-T. Ko, J.-H. Park, W. Ratcliff II, P. A. Sharma, N. Harrison, Y. J. Choi, Y. Horibe, S. Lee, S. Park, H. T. Yi, C. L. Zhang, and S.-W. Cheong, Phys. Rev. Lett. **101**, 137203 (2008).

[17]K. T. Ko, H. J. Noh, J. Y. Kim, B.-G. Park, J.-H. Park, A. Tanaka, S. B. Kim, C. L. Zhang, and S-W. Cheong Phys. Rev. Lett. **103**, 207202 (2009).

[18]Y. Tang, Y. Sun, and Z. Cheng, J. Appl. Phys. **100**, 023914 (2006).

[19]M. Patra, S. Majumdar, and S. Giri, J. Phys.: Condens. Matter **22**, 116001 (2010).

[20]C. Schlenker, S. S. P. Parkin, J. C. Scott, and K. Howard, J. Magn. Magn. Mater. **54**, 801 (1986).

[21]A. Hochstrat, C. Binek, W. Kleemann, Phys. Rev. B **66**, 092409 (2002).

[22]C. Binek, Phys. Rev. B **70**, 014421 (2004).

[23]A. Hoffmann, Phys. Rev. Lett. **93**, 097203 (2004).

[24]D. Paccard, C. Schlenker, O. Massenet, R. Montmory, and A. Yelon, Phys. Status Solidi **16**, 301 (1966).

[25]U. Nowak, K. D. Usadel, J. Keller, P. Miltényi, B. Beschoten, and G. Güntherodt, Phys. Rev. B **66**, 014430 (2002).

**FIG. 1.** (Color online). (a) Schematic diagram of the spin configuration of a single-phase magnet with two magnetic sublattices. The spin reversal of the FM sublattice feels the pinning from the AFM sublattice. (b) Crystal structure of $LuFe_2O_4$ and $YbFe_2O_4$.

**FIG. 2.** (Color online). Temperature dependence of magnetization of $YbFe_2O_4$ measured in the FC mode with $H$=1 kOe. The magnetic transition at $T_1$=245 K corresponds to the ferrimagnetic ordering of $Fe^{2+}/Fe^{3+}$ sublattice and the transition at $T_2 \approx 50$ K is due to the coupling between $Yb^{3+}$ and $Fe^{2+}/Fe^{3+}$ sublattices. The inset shows the $M$-$T$ curves in the FC mode with $H$=5 and 13 T.

**FIG. 3.** (Color online). The $M$-$H$ hysteresis loops of $YbFe_2O_4$ at 5 K with both the ZFC and FC modes. For comparison, the inset shows the FC hysteresis loop of $LuFe_2O_4$ measured at 5 K with a cooling field of 13 T.

**FIG. 4.** (Color online). Cooling magnetic field dependence of $H_{EB}$ of $YbFe_2O_4$ at 5 K. Inset: enlarged view of the $M$-$H$ loops with different $H_{cool}$=0, 5, 10, and 50 kOe.

**FIG. 5** (Color online). Temperature dependence of $H_{EB}$ measured with a cooling field of 13 T. The insets show the $M$-$H$ loops and the cooling field dependence of EB at 25 K.

**FIG. 6** (Color online). Training effect of EB at 5 K. The main panel shows the consecutive hysteresis loops measured after FC in 13 T. For clarity, only the loops with cycling number $n$=1, 2, 3, and 10 are plotted. The inset shows $H_{EB}$ vs $n$. The solid line is the fitting curve with Eq. (2).

Fig. 1

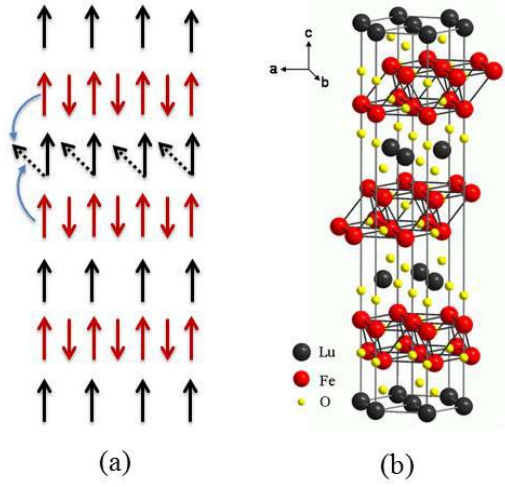

(a)      (b)

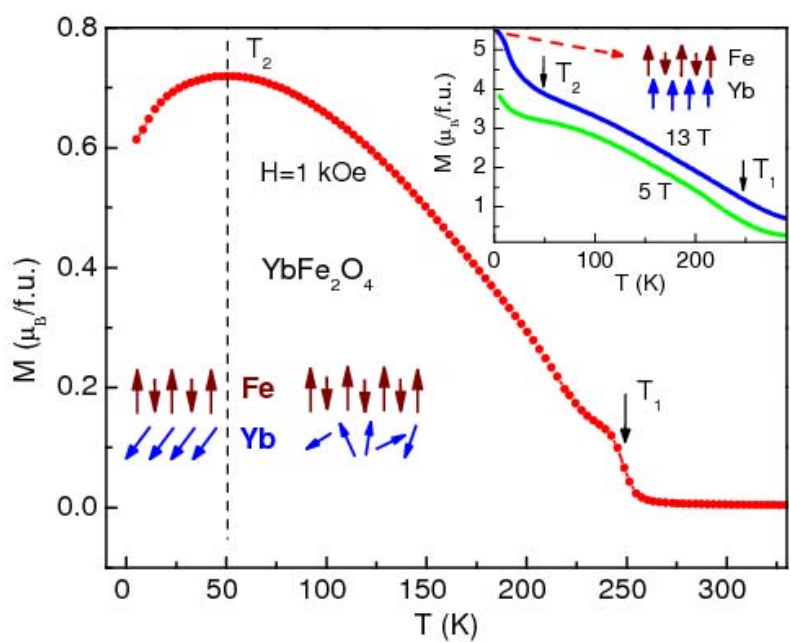

Fig. 3

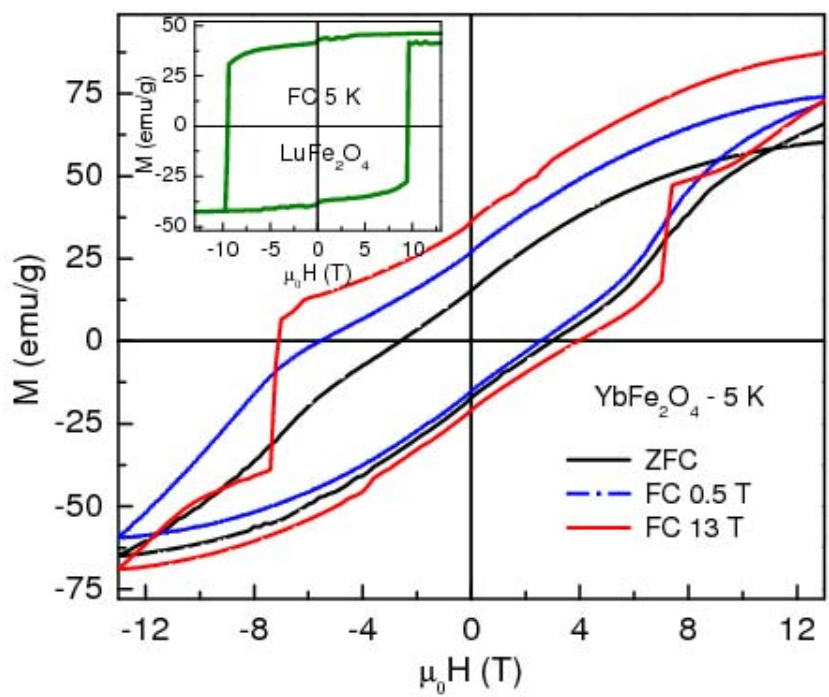

Fig. 4

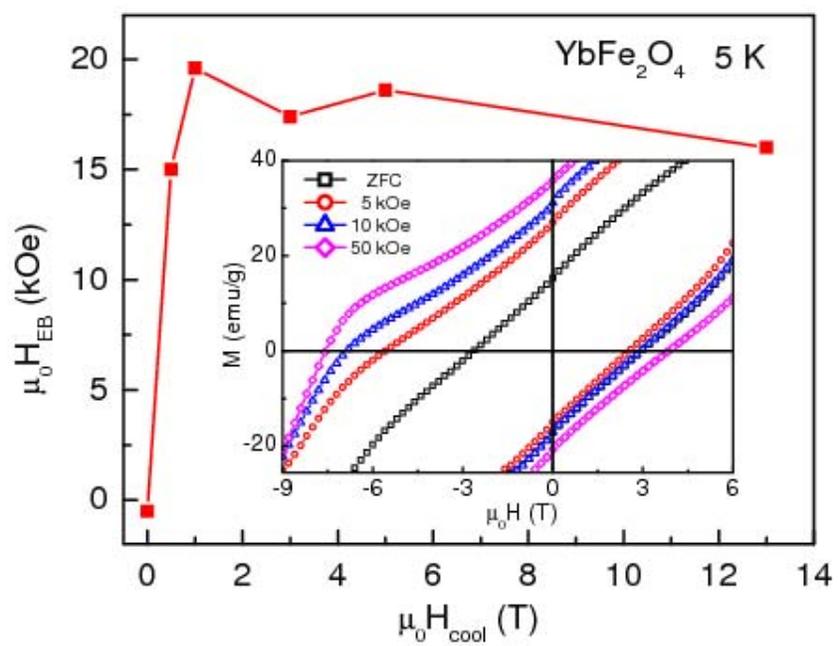

Fig. 5

Fig. 6

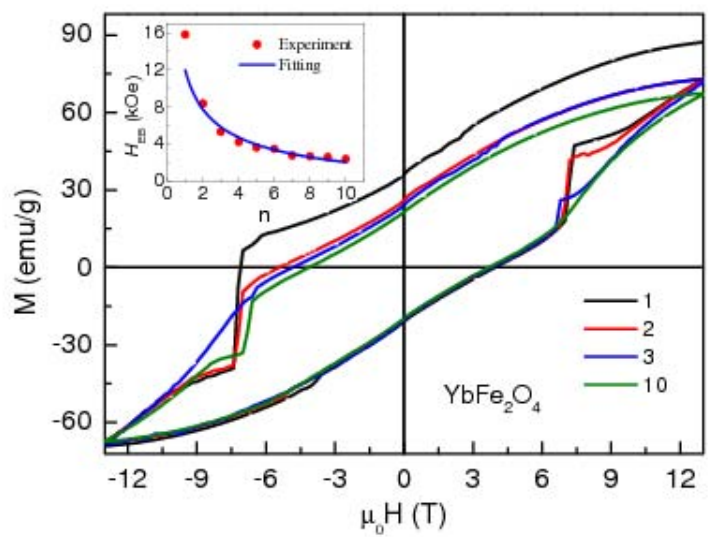